\documentstyle[aps,prl,twocolumn]{revtex}

\begin{document}
\title{Nonadiabatic Approach to Spin-Peierls Transitions via Flow Equations}

\author{G\"otz S.~Uhrig}
\address{Institut f\"ur Theoretische Physik, Universit\"at zu K\"oln,
 Z\"ulpicherstra\ss e 77, 50937 K\"oln, Germany.}
\draft
\maketitle
\begin{abstract}
The validity of the adiabatic approach to spin-Peierls
transitions is assessed. An alternative approach is 
developed which maps the initial magneto-elastic problem to an
effective magnetic problem only. Thus the equivalence
of magneto-elastic solitons and magnetic spinons is shown.
No soft phonon is required for the transition. Temperature
dependent couplings are predicted in accordance with the ana\-lysis
of experimental data.
\end{abstract}

Around 1980 there was a great interest in the phenomenon of 
spin-Peierls (SP) transitions \cite{bray83}
where the coupling of lattice degrees of freedom to 
quasi one-dimensional ($d=1$) magnetic degrees of freedom leads
to a phase transition into a dimerized phase.
This interest has been vividly renewed recently due to the discovery
of the first inorganic SP substance CuGeO$_3$ \cite{hase93a,bouch96}.

 The instability of the coupled spin-phonon system
towards dimerization results from the susceptibility of 
the magnetic $d=1$ subsystem towards $2k_{\rm F}$
perturbations, i.e.\ dimerization.
The gain in magnetic energy due to dimerization
overcompensates the loss in elastic energy of the lattice distortion.
The present work is motivated by the wealth of information
available for the SP transition in CuGeO$_3$.

The theoretical picture of SP transitions has been
developped in a number of articles (e.g.\ 
\cite{pytte74b,bray75,cross79,cross79b,nakan80nakan81,buzdi83b,fujit84fujit88}).
It  relies so far mostly on an {\em adiabatic} treatment of the phonons.
Cross and Fisher discussed this point most comprehensively \cite{cross79}.
By random phase approximation (RPA) they investigate the stability of 
the uniform phase. The RPA 
is the consistent extension of the mean-field treatment 
on the one-particle level to the two-particle level of
susceptibilities.
Cross and Fisher's point  is the importance of a
``{\em pre-existing soft phonon}''. This means that the phonons responsible
for the distortion have a {\em low} energy already {\em before} the 
interaction with the correlated system is taken into account.
Moreover, they require that the phonon dispersion
perpendicular to the chain direction is very large. Thus
the lattice prefers that whole planes perpendicular to the 
correlated chains move coherently so that the moving objects are heavy.

The reason for the above requirements is that the mean-field approach is 
appropriate if the fluctuations are small compared to the 
expectation value. This is the case if the distortions are made
up by a large number of phonons which in turn means that the
phonon energies must be small. The phonons must be slow
and heavy. Then it is plausible to consider the phonon subsystem
as the slow subsystem which is renormalized by the fast magnetic degrees of
freedom. In this picture, the SP transition is signalled by the
vanishing of a renormalized phonon frequency \cite{cross79}.

Whereas the experimental data for organic SP substances
such as (TTF)CuS$_4$C$_4$(CF$_3$)$_4$
 supports the RPA approach \cite{cross79}, the experimental
evidence for the inorganic CuGeO$_3$ points into the opposite
direction \cite{brade96a}. Braden {\it et al}.\ found that 
two out of four optical phonons allowed by symmetry contribute
 appreciably to the  distortion in a ratio of 3:2 \cite{brade96a}.
The more important phonon is at 6.8Thz (330K);
its dispersion at the zone boundary is essentially flat decreasing towards
the zone centre  to $\approx 3$THz.
The other phonon is at 3.2THz (150K) and practically dispersionsless.
The magnetic exchange coupling $J$ lies in the range 115K to 160K
\cite{nishi94,riera95,casti95,uhrig97a,fabri97a}. It is evident that the
assumption of a pre-existing soft phonon is  inadequate.
Things get even worse if one takes into consideration the 
 results for a XY chain with $d=1$, Einstein phonons \cite{caron96}
which show that the phonon energy $\omega$ must be small compared
to the resulting gap $\Delta$ for the adiabatic approach
to be reasonable. The condition $\omega<\Delta$, however, is
 definitely not fulfilled in CuGeO$_3$ with
$\Delta=23K$ \cite{regna96a}. In view of these facts it is not
astounding that so far no phonon softening at the SP
transition was found experimentally.
For comparison, we recall the numbers for the
best-known substance (TTF)CuS$_4$C$_4$(CF$_3$)$_4$ with soft phonon:
$J=68$K, 
$T_{\rm SP}=12$K ($\Rightarrow \Delta \approx 1.77T_{\rm SP}
\approx 20$K), $\omega \approx 10$K 
where $\omega < \Delta$ is fulfilled  \cite{bray83}.

In the present work we develop a route to Peierls
transitions  not based on the assumption of phononic adiabaticity.
 Results of previous works
\cite{pytte74a,brand74b,essle97} on nonadiabaticity will be extended.
We will view the phonon subsystem as fast and the
spin subsystem as slow.
The unperturbed spin system is always gapless,
i.e.\ the magnetic subsystem always  has low-lying excitations 
well below any (optical) phonon energy. These low-lying 
excitations are influenced most by the interaction of phonons
and spins. To them the phonons are  fast. So we 
treat  the phonons as quickly adapting  and derive
an effective dressed spin model.
Pytte did the same for an Ising model which allowed
the rigorous elimination of phonons \cite{pytte74a} stressing already
the importance of avoiding a mean-field approximation for the 
displacements.

Technically we use the recently developed flow equation approach
to treat the spin-phonon system \cite{wegne94}. The  idea
is to rotate away the direct interaction with phonons  similar to what
is done in Fr\"ohlich's approach \cite{frohl52}. This approach has been
improved considerably by Lenz and Wegner for electron-phonon
interactions \cite{lenz96}. In the improved version the generated
effective couplings are by far less singular than in Fr\"ohlich's
approach.

The flow equation approach ``diagonalizes'' a Hamiltonian
in a continuous unitary transformation parametrized by $l\in [0,\infty]$.
This means $H(0)$ is the bare Hamiltonian as given and $H(\infty)$
is the resulting (more) diagonal Hamiltonian. The unitary transformation
is defined by its antihermitean infinitesimal generator $\eta(l)$
via
\begin{equation}
\label{hfluss}
\frac{dH}{dl} = [\eta(l), H(l)] \ .
\end{equation} 
A good choice for $\eta$ is 
\begin{equation}
\label{etafluss}
\eta = [H_{\rm D}(l), H(l)]
\end{equation}
where $H_{\rm D}$ is the  
suitably chosen diagonal Hamiltonian \cite{wegne94}. The main feature of
$\eta$ as defined in (\ref{etafluss}) is that it respects the idea of
renormalization in that it eliminates first matrix elements connecting
very different energies \cite{kehre96b}. 

The SP system we consider is given by the Hamiltonian
\begin{mathletters}
\label{hdef}
\begin{eqnarray}
\label{hdefa}
H &=& H_{\rm S} + H_{\rm B} + H_{\rm SB}\qquad 
H_{\rm S} = \sum_{\vec{q}} d(\vec{q}) \vec{S}_{\vec{q}}\vec{S}_{-\vec{q}}
\\
H_{\rm B}& =& \sum_{\vec{q}} \omega(\vec{q}) b^+_{\vec{q}}
b^{\phantom +}_{\vec{q}}
\qquad
 H_{\rm SB} = \sum_{\vec{q}} 
A_{\vec{q}} (b^+_{\vec{q}}+b^{\phantom +}_{-\vec{q}})  \\
A_{\vec{q}}  &=& \sum_{\vec{k}} g(\vec{q},\vec{k}) 
\vec{S}_{\vec{k}}\vec{S}_{-\vec{k}-\vec{q}} \ .
\end{eqnarray}
\end{mathletters}
in obvious notation in momentum space. Note that according to
\cite{wegne94} $A_{\vec{q}}$ should be normal-ordered
$A_{\vec{q}} \to A_{\vec{q}}- \langle A_{\vec{q}}\rangle$.
The particular choice (before eq.\ \ref{fin3} and eq.\ \ref{AA})
  for $A_{\vec{q}}$ will circumvent this problem.
We assume inversion symmetry
so that $\omega(\vec{q}) = \omega(-\vec{q})$ and 
$d(\vec{q}) = d(-\vec{q})$.
Hermiticity requires $A^+_{-\vec{q}}= A^{\phantom +}_{\vec{q}}$ or
equivalently $g^*(-\vec{q},-\vec{k})= g(\vec{q},\vec{k})$.
The linear boson terms become
$l$-dependent for the unitary transformation
\begin{equation}
H_{\rm SB}(l) = \sum_{\vec{q}} \left(T^{\phantom +}_{\vec{q}}(l) b^+_{\vec{q}}
+ T^+_{\vec{q}}(l) b^{\phantom +}_{\vec{q}} \right) 
\end{equation}
with the starting condition $T_{\vec{q}}(0)=A_{\vec{q}}$.
The objective of the unitary transformation is to disentangle
phonons and spins. Thus we choose $H_{\rm D} = H_{\rm S} + H_{\rm B}$.
To leading order in $g$
we do not need to consider a possible $l$-dependence of $H_{\rm D}$
since the $l$-dependent terms enter only in order $g^2$ as we will see.
These induced terms of order $g^2$ and higher
 lead to a new contribution $\Delta H$ to the Hamiltonian.
Introducing the Liouville operator ${\cal L}$ for the commutation
with $H_{\rm S}$: ${\cal L}A:=[H_{\rm S},A]$ we choose for the
generator $\eta$
\begin{mathletters}
\begin{eqnarray}
&&\eta(l) = [H_{\rm D},H_{\rm SB}]\\
&&\; = \sum_{\vec{q}} \left(({\cal L}+\omega(\vec{q}))
T^{\phantom +}_{\vec{q}}(l) b^+_{\vec{q}}
+ ({\cal L}-\omega(\vec{q})) T^+_{\vec{q}}(l) b^{\phantom +}_{\vec{q}} \right)
 \label{etadef}
\end{eqnarray}
\end{mathletters}
which is motivated by (\ref{etafluss}).
The flow equation (\ref{hfluss}) leads to
\begin{eqnarray}
&&\frac{dH}{dl} = [\eta,H_{\rm SB}] + [\eta,\Delta H] \nonumber \\ 
&& - \sum_{\vec{q}} \left(({\cal L}+\omega(\vec{q}))^2
T^{\phantom +}_{\vec{q}}(l) b^+_{\vec{q}}
+ ({\cal L}-\omega(\vec{q}))^2 T^+_{\vec{q}}(l) b^{\phantom +}_{\vec{q}}\right)
\label{efluss}
\end{eqnarray}
In linear order in $g$ we have to fulfil the flow equation
\begin{equation}
\frac{dT_{\vec{q}}}{dl} = -({\cal L} + \omega(\vec{q}))^2 T_{\vec{q}}
\end{equation}
which is formally solved by
\begin{equation}
\label{sfluss}
T_{\vec{q}}(l) = \exp\left(-({\cal L} + \omega(\vec{q}))^2 l \right) 
A_{\vec{q}}\ .
\end{equation}
Based on (\ref{sfluss}) the additional Hamilton part $\Delta H$ 
 can be calculated 
\begin{mathletters}
\label{newham}
\begin{eqnarray}
&& \frac{d\Delta H}{dl} = [\eta,H_{\rm SB}] + {\cal O}(g^3)
\\
&&\; = -\sum_{\vec{q},\vec{k}} 
\left[T^{\phantom +}_{\vec{q}} b^+_{\vec{q}} + 
T_{\vec{q}} b^{\phantom +}_{\vec{q}},
D^{\phantom +}_{\vec{q}} b^+_{\vec{q}} -
D^+_{\vec{q}} b^{\phantom +}_{\vec{q}}
\right] + {\cal O}(g^3)\\
&&\; = -\sum_{\vec{q}}\left(D^+_{\vec{q}} T^{\phantom +}_{\vec{q}} +
T^{+}_{\vec{q}}D^{\phantom +}_{\vec{q}}  \right)\\
&&\quad +  \label{plusplus}
\sum_{\vec{q},\vec{k}}\left(b^+_{\vec{q}}b^+_{\vec{k}}
[ D^{\phantom +}_{\vec{q}},T^{\phantom +}_{\vec{k}}] + \, \mbox{h.c.} \right)
\\
&&\quad + 
\sum_{\vec{q},\vec{k}}\left(b^+_{\vec{q}}b^{\phantom +}_{\vec{k}}
\left( [ T^{\phantom +}_{\vec{q}},D^{+}_{\vec{k}}] +  
[ D^{\phantom +}_{\vec{q}},T^{+}_{\vec{k}}]\right) \right)+ {\cal O}(g^3)
\end{eqnarray}
\end{mathletters}
where we used the shorthand $D_{\vec{q}}=
({\cal L} + \omega({\vec{q}}))T_{\vec{q}}$.

To obtain from
(\ref{newham}) an effective spin Hamiltonian we use
a mean-field approach and replace the quadratic boson terms
by its expectation values. This is absolutely systematic in the sense of an
expansion in $g$. Taking 
the expectation values neglects fluctuation effects of the order
$g^2$ due to the interaction. But since the two-boson terms
appear only as $g^2$-terms the total error due to the mean-field
treatment is of the order $g^4$. Applying the same mean-field
approach to the unspecified $g^3$ terms annihilates them 
because they contain necessarily an odd number of boson
operators. Thus the effective spin model is exact up to 
${\cal O}(g^4)$.

Replacing $b^+_{\vec{q}} b^{\phantom +}_{\vec{k}}$ by
$\delta_{\vec{q},\vec{k}} (\exp(\omega(\vec{q})/T)-1)^{-1}$
and omitting the terms in (\ref{plusplus}) we obtain
finally
\begin{mathletters}
\label{fin1}
\begin{eqnarray}
&&\frac{d\Delta H}{dl} =
 \sum_{\vec{q}} \left( X_{\vec{q}} + 
\coth\left(\frac{\omega(\vec{q})}{2T}\right)  Y_{\vec{q}}\right)\quad
\mbox{with}\\
&&X_{\vec{q}} = -\frac{1}{2}\left(
 D^{\phantom +}_{\vec{q}}T^{+}_{\vec{q}}
+  D^{+}_{\vec{q}}T^{\phantom +}_{\vec{q}}
+T^{\phantom +}_{\vec{q}}D^{+}_{\vec{q}}
+  T^{+}_{\vec{q}}D^{\phantom +}_{\vec{q}}\right)
\\
&&Y_{\vec{q}} = \frac{1}{2}  
\left(
[ T^{\phantom +}_{\vec{q}},D^{+}_{\vec{q}}] +  
[ D^{\phantom +}_{\vec{q}},T^{+}_{\vec{q}}]
\right)\ .
\end{eqnarray}
\end{mathletters}
From (\ref{fin1}) the Hamiltonian corrections  $\Delta H_{X/Y}$
are found by integration over $l$ and summation over $\vec{q}$:
$\Delta H_X = \int_0^\infty \sum_{\vec{q}} X_{\vec{q}}  dl$ and
$\Delta H_Y = \int_0^\infty \sum_{\vec{q}} 
\coth\left({\omega(\vec{q})}/(2T)\right) Y_{\vec{q}}  dl$.

In  order to get an impression of what (\ref{fin1}) means
we assume $J\ll \omega(\vec{q})$ and calculate the leading
contributions in ${\cal L}$ to $X_{\vec{q}}$ (even in ${\cal L}$) and
$Y_{\vec{q}}$ (odd in ${\cal L}$).
After some algebra and integration we find
\begin{mathletters}
\label{fin2}
\begin{eqnarray}
\Delta H_X &=&
-\sum_{\vec{q}} 
\frac{1}{\omega(\vec{q})} A^+_{\vec{q}}A^{\phantom +}_{\vec{q}} =
 \frac{-1}{\omega}\sum_{i} 
 A^+_{i}A^{\phantom +}_{i}\label{fin2b} \\
\Delta H_Y &=&
\frac{1}{2}\sum_{\vec{q}} 
\frac{1}{\omega^2(\vec{q})}
\coth\left(\frac{\omega(\vec{q})}{2T}\right)
\left[ A^+_{\vec{q}}, {\cal L}A^{\phantom +}_{\vec{q}}\right] \\
&=&
\frac{1}{2\omega^2}\coth\left(\frac{\omega}{2T}\right) \sum_{i} 
\left[ A^+_{i}, {\cal L}A^{\phantom +}_{i}\right]\label{fin2d}
\end{eqnarray}
\end{mathletters}
where we simplified the formulae (\ref{fin2b}) and (\ref{fin2d})
in real space one step further approximating the phonons by Einstein phonons.
The term $\Delta H_X$ corresponds to the results obtained previously
by other methods \cite{pytte74a,brand74b,essle97}. 
To the author's
knowledge, the $T$-dependent term $\Delta H_Y$ has not yet been described.
The result of Pytte \cite{pytte74a} is found back by observing that
in the Ising model ${\cal L} A_i$ vanishes since all terms involving
only $S^z$ commute, i.e.\ $\Delta H_Y$ becomes zero. The result in
\cite{essle97} is retrieved on observing that $g^2/\omega$ is proportional
to $J^2/(m\omega^2)$ in \cite{essle97} since the displacments equal
$u_i = (b_i+b^\dagger_i)/\sqrt{2m\omega}$. Neglecting the phononic
kinetic energy while keeping their potential one constant corresponds
to the limit $m\to0$ with $m\omega^2$ constant, i.e.\ 
$\omega\to\infty$. So the $g^2/\omega$
term is constant and kept while terms  $g^2 {\cal O}(\omega^{-2})$ like
$\Delta H_Y$ are neglected in \cite{essle97}.
To further enhance the plausibility of the result (\ref{fin2})
we note that it equals the result one  gets by
Fr\"ohlich's method \cite{frohl52} in the two leading orders $g^2/\omega$
and $g^2J/\omega^2$.
The difference of the flow equation approach and Fr\"ohlich's approach
appears only in the $1/\omega^3$ terms coming from (\ref{fin1}). Insofar
Fr\"ohlich's approach can also be used to derive (\ref{fin2}). The flow 
equation approach, however, is a better starting point for future 
higher-order calculations in $g/\omega$ and $J/\omega$ which take 
$\ell$-dependent couplings into account. This is the reason why this
method is chosen here.

Specifically, we consider first strictly one dimensional phonons
 $A_i = g (\vec{S}_i\vec{S}_{i+1}- \vec{S}_i\vec{S}_{i-1})$.
This choice guarantees $\langle A_i \rangle =0$ in the 
symmetry unbroken phase so that $A_i$ is normal-ordered.
With this $A_i$ we have 
\begin{eqnarray}\label{fin3}
\Delta H_X &=&  \frac{g^2}{\omega} \sum_{i} 
(\vec{S}_i \vec{S}_{i+1}+
\frac{1}{2}\vec{S}_{i} \vec{S}_{i+2}-\frac{3}{8} )\ .
\end{eqnarray}
For $\Delta H_Y$ we have to know $d(\vec{q})$ in (\ref{hdefa}).
We assume nearest and next-nearest neighbour interaction $J$ and
$\alpha J$, respectively: $d(\vec{q})= J(\cos(q_1)+\alpha\cos(2q_1))$.
One obtains
\begin{eqnarray}\label{fin4}
&&\Delta H_Y =  \frac{J}{4}\frac{g^2}{\omega^2}
\coth\left(\frac{\omega}{2T}\right)\times\\
&&\; \sum_{i} 
(- (3-3\alpha)\vec{S}_i\vec{S}_{i+1} 
+(3-5\alpha)\vec{S}_i\vec{S}_{i+2}
+ 2\alpha \vec{S}_i\vec{S}_{i+3}) \nonumber
\end{eqnarray}
where products with four different spins are omitted.

Even if no frustration is present in the original
model ($\alpha=0$) the dressing of the spins with phonons
induces  frustration $\alpha_{\rm eff}>0$.
 The couplings are
temperature dependent since they are mediated by the phonons.
Using the term ``spinon'' for a purely magnetic elementary $S=1/2$ excitation
and the term ``soliton'' for the joint magnetic and elastic
$S=1/2$ excitation we state that the solitons of the Hamiltonian
(\ref{hdef}) are unitarily equivalent to the spinons of the Hamiltonian
$H=H_{\rm S} + H_{\rm B} + \Delta H_X +\Delta H_Y$. This shows that
solitons and spinons are in essence the same entity and puts 
Affleck's supposition in this respect \cite{affle97} on a quantitative basis.

The low-lying excitations of the frustrated Heisenberg chain
are spinons which are gapless for $\alpha_{\rm eff}< \alpha_{\rm c}=
0.241$ \cite{julli83okamo92} and gapful above $\alpha_{\rm c}$
\cite{chitr95white96} where the system undergoes also a
spontaneous symmetry breaking of the translational symmetry towards
a dimerized phase.
The continuum starts right at the gap energy \cite{shast81}.
These facts imply that a {\em single} chain shows a SP
transition only above a certain value of the interaction in 
contrast to the results of the adiabatic treatment (see also
\cite{caron96}). Furthermore, no ``double gap'' feature \cite{ain97}
occurs in a {\em single} chain.

An elastic interchain coupling in a chain ensemble
is included if the local phonons influence also neighbouring chains
\begin{eqnarray}\nonumber
A_{i,j} &=& g (\vec{S}_{i,j}\vec{S}_{i+1,j}- \vec{S}_{i,j}\vec{S}_{i-1,j}+ \\
&&
f\sum_{<j,j'>}(\vec{S}_{i,j'}\vec{S}_{i+1,j'}-
\vec{S}_{i,j'}\vec{S}_{i-1,j'})
\label{AA}
\end{eqnarray}
where $j$ is the chain index and $j$ and $j'$ are adjacent chains. The
factor $|f|<1$ indicates the influence of a certain distortion on one
chain onto adjacent chains. 
Due to the commutators in (\ref{fin2d}) a finite $f$  changes 
$\Delta H_Y$ in (\ref{fin4}) only by renormalizing 
$g^2 \to \tilde g^2 = g^2(1+zf^2)$ where
$z$ is the number of neighbouring chains each chain has.
The same renormalization takes place in $\Delta H_X$. Additionally,
terms linking different chains occur like
$-(g^2 f/\omega)(\vec{S}_{i,j}\vec{S}_{i+1,j}- \vec{S}_i\vec{S}_{i-1}) 
(\vec{S}_{i,j'}\vec{S}_{i-1,j'}- \vec{S}_{i,j'}\vec{S}_{i-1,j'})$. 
These terms drive really the finite temperature SP transition
since they enable at low enough temperature 
a coherent dimerization throughout the whole lattice.
 We will call these terms coherence terms. 
Their influence on the low-lying excitations is to confine pairs
of spinons (solitons) to triplets or to singlets.
 For the realistic case of weak coupling
$g^2 f/\omega \ll J$ a mean-field treatment is justified. This amounts up
to the treatment of dimerized, frustrated chains with  self-consistently
determined dimerization. Hence the confinement is the same
as in dimerized chains (see e.g.\ \cite{affle97,uhrig96b}).
This explains why the adiabatic approaches based on dimerized, frustrated
chains capture correctly the physics of the dimerized SP phase at low $T$.

The main difference of the fast-phonon scenario to the adiabatic one
is the absence of
a soft phonon at the transition. The transition is characterized by
growing  domains of coherent dimerization the size of which
diverges at $T_{\rm SP}$. No renormalized phonon frequency needs to
vanish. Interestingly, the RPA shows similar results in the nonadiabatic
parameter regime \cite{schulz}.\\
How does this fit to the approach used so far?
In the usual RPA treatment \cite{cross79} the
phononic self-energy contributes not only a real part but also 
an equally strong imaginary part which stands for strong damping
\cite{note}.
Thus the real, untransformed
 phonons are not appropriate quasi-particles.
If we transform the  propagator of the real phonons in the
same unitary way as the Hamiltonian we see that via
$b_i^\dagger \to b_i^\dagger - \frac{g}{\omega}A_i + \ldots$
not only the transformed phonons matter but also $S=0$ excitations
of the effective spin model. This means that the observation
of the real phonons  reveals not only a sharp peak at the high
frequency $\omega$ but also a  continuum at low
frequencies of the order of $J$. The low energy continuum changes
on passing through the spin-Peierls transition. In the vicinity
of $T_{\rm SP}$ one expects some critical fluctuations close to
zero energy. Below
$T_{\rm SP}$ a gap should appear which equals  twice 
the triplet gap  or less if bound states are present 
 \cite{uhrig96b,bouze97a}. But the 
phonon peak is not lowered towards zero energy at the transition.
This is in accordance with the results known so far for CuGeO$_3$.

We attempt to estimate orders of magnitude for the couplings
in CuGeO$_3$. Let us assume that $T_{\rm SP}$ is of the order
of $g^2/\omega$ then $g^2/\omega^2\approx T_{\rm SP}/\omega\approx
15K/150K=0.1$ is roughly one tenth which
justifies the expansion in $g^2/\omega^2$.

The extension of (\ref{fin2}) to several phonons is straightforward.
Using the phonon energies and their relative distortion as experimental input
and the coupling $g$ as fit parameter it is possible to reproduce
the experimental $\chi(T)$ data nearly as well as in \cite{fabri97a}
from where also the $\chi(T)$ data was taken.
This shows that the assumption of $T$-dependent couplings is {\em not}
ruled out by the $\chi(T)$ data. With the same parameter as for the
$\chi(T)$ fit we find $J(50\mbox{K})=162K$ and $J(300\mbox{K})=140K$ which 
agrees very well to $J(50\mbox{K})=158K$ and $J(300\mbox{K})=136K$ deduced 
by Fabricius and L\"ow \cite{fabri97b} from experimental
$S(q,\omega)$ data. This excellent agreement confirms the validity
of the approach used and in particular the prediction of $T$ dependent
couplings.

In summary, we discussed the validity of the phonon adiabatic
approach for SP transitions and in particular CuGeO$_3$. The phonon adiabatic
approach is inadequate for the latter system. We developed a promising
alternative approach relying on the flow equation technique. 
Magneto-elastic solitons are mapped to 
magnetic spinons in an effective, purely magnetic Hamiltonian.
The phonon dynamics induces a $T$-dependent frustration. No soft phonon
signals the SP transition which is driven by the coherence terms in 
an effective magnetic model.

The author is indebted to M. Braden  for making his results
available to him prior to publication. He acknowledges
helpful discussions with  M. Braden, W. Brenig, B. B\"uchner,
E. M\"uller-Hartmann, and H.~J. Schulz. This work was supported by the DFG
through SFB 341.


\end{document}